# Unraveling the multistage phase transformations in monolayer Mo-Te compounds

Zemin Pan,[1] Tao Jian,[1] Hui Zhang,[1] Xiaoyu Lin,[1] Chao Zhu,[1] Jinghao Deng,[1] Zhengbo Cheng,[1] Chuansheng Liu,[1,a)] and Chendong Zhang[1,a)]

**Affiliations:**

[1]School of Physics and Technology, Wuhan University; Wuhan 430072, China.

a)Authors to whom correspondence should be addressed: csliuan@whu.edu.cn and cdzhang@whu.edu.cn

**Abstract:** Monolayer $MoTe_2$ exhibits a variety of derivative structural phases and associated novel electronic properties that enable a wealth of potential applications in future electronic and optoelectronic devices. However, a comprehensive study focusing on the complexities of the controllable phase evolution in this atomically thin film has yet to be performed. This work aims to address this issue by systematically investigating molecular beam epitaxial (MBE) growth of monolayer Mo-Te compounds on bilayer graphene substrates. By utilizing scanning tunnelling microscopy (STM), we explored a series of thermally driven structural phase evolutions including distinct T'-$MoTe_2$, H-$MoTe_2$, $Mo_6Te_6$ nanowires, and multistoichiometric $MoTe_{2-x}$. Furthermore, we carefully investigated the critical effects of the growth parameters—annealing temperature and time and tellurium concentration—on the controllable and reversible phase transformation within monolayer $MoTe_{2-x}$. The findings have significant implications for understanding the thin film synthesis and phase transformation engineering inherent to two-dimensional crystals, which can foster further development of high-performance devices.

**Main Text:** Monolayer transition metal dichalcogenides (TMDs) have garnered extensive interest owing to their exotic electrical and optoelectronic properties.[1-3] TMD materials have shown great diversity, ranging from semimetallic to semiconductor, superconductor, and even insulator states.[4,5] One of their key advantages lies in their abundance of different structural phases, such as hexagonal 2H, octahedral 1T, monoclinic or distorted octahedral 1T', and orthorhombic $T_d$ structures, and the electronic band structure and emergent phenomena are closely related to these structural phases.[6-10] Very recently, one-dimensional (1D) counterparts of monolayer TMDs, *i.e.*, transition metal monochalcogenide wires with single-unit-cell widths, were reported for Mo-Te, Mo-Se, and W-Te compounds, which can exhibit appealing 1D charge density wave (CDW) states and quantum-confined Luttinger liquid behavior.[11-13] These findings further extend the phase control studies of M-X compounds to the areas of stoichiometry and dimensional engineering.

Within the TMD family, monolayer (ML) or few-layer $MoTe_2$ exhibits distinctive properties in many respects. For instance, few-layer or ML structures of $T_d$-$MoTe_2$ were reported to manifest ferroelectric and superconductivity,[14] while 1T'-$MoTe_2$ was experimentally identified as a topological Weyl semimetal.[15] The most recent breakthrough is the direct observation of the fractional quantum effect in twisted $MoTe_2$.[16] From the perspective of structural phase engineering, Mo-Te is expected to have the most diverse structural phases due to the small difference in the formation energy among the common TMD components.[6,17] Indeed, several separate studies recently reported numerous distinct Mo-Te structural phases, including various mirror twin boundary (MTB) superstructures in H-MoTe$_2$,[18-21] Mo$_5$Te$_8$,[22,23] and Mo$_6$Te$_6$.[11,24] These phases can dramatically alter the lattice geometries and electron–electron interactions or even generate new superlattice potentials, leading to remarkable phase-dependent properties such as coloring-triangle flat bands, CDWs, and Tomonaga–Luttinger liquid behavior.[11,19-22] While exciting experimental identifications and physical characterizations of each structural phase of ML Mo-Te compounds continue to emerge, the systematic control of the synthesis and transformation within this enriched phase diagram has thus far been largely unexplored.

In this work, we reveal the controllable thermally driven structural phase evolution in a Mo-Te ML by combining scanning tunnelling microscopy (STM) and molecular beam epitaxy (MBE). By controlling the growth substrate temperature (ranging from 150°C to 550°C), we sequentially obtain the low-temperature phase T'-MoTe$_2$, H-MoTe$_2$, MoTe$_{2-x}$ ($x$ ranging from 0 to 0.4), and high-temperature phase 1D Mo$_6$Te$_6$ nanowires (NWs). Through refined control of the postgrowth annealing process, we thoroughly investigate the refined stoichiometry-dependent structural transformation in ML MoTe$_{2-x}$, which can be considered to share the same building block, namely, an MTB. We found that the $x$ value increases as the annealing temperature increases, accompanied by a change in the MTB triangular loops from local defects to global superstructures. Two distinct phases are acquired at 450 °C and 500 °C, corresponding to $x = 0.3$ and $x = 0.4$, respectively. The $x = 0.3$ phase can be effectively isolated from other phases in the growth parameter space[21], whereas the $x = 0.4$ phase (*i.e.*, Mo$_5$Te$_8$)[22,23] is always mixed with other phases ($x = 0.3$ and Mo$_6$Te$_6$) and shows limited coverage. Moreover, in our experiments, a controlled reversible transformation from Mo$_5$Te$_8$ ($x$ = 0.4) to nearly pristine MoTe$_2$ ($x$ = 0) is achieved. Our findings not only provide a systematic strategy for manipulating multistoichiometric phases in ML Mo-Te compounds but also support the phase engineering inherent in these materials.

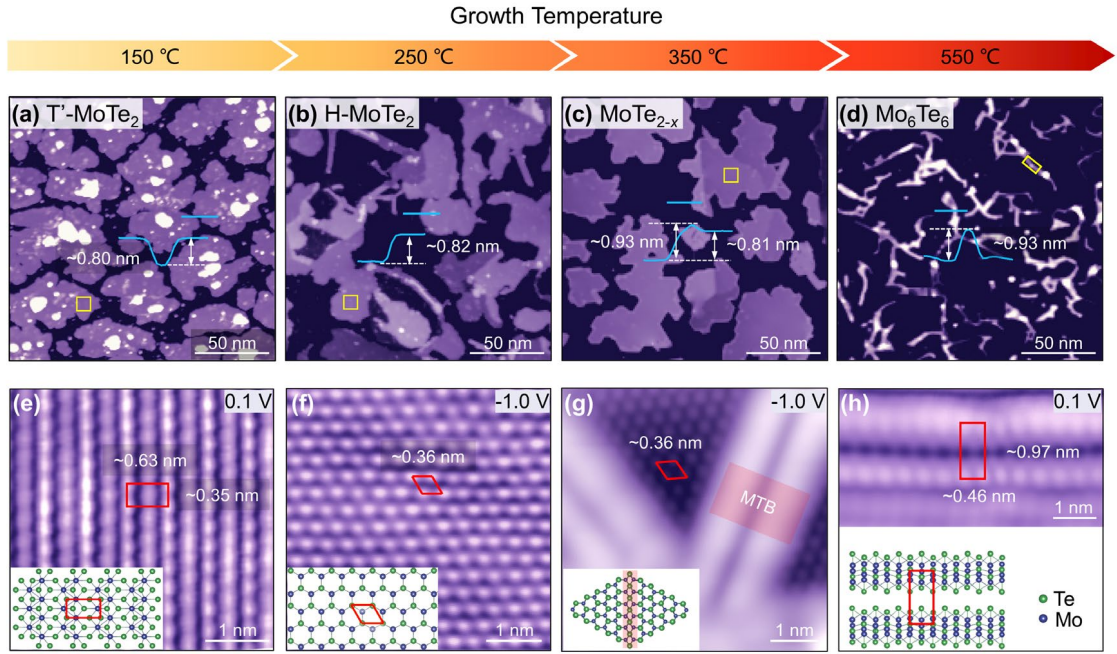

**Fig. 1.** Phase control of ML Mo-Te compounds grown on BLG/SiC. (a)-(d) Large-scale STM topographic images of samples grown at different substrate temperatures (during deposition); the corresponding line profiles were taken along the blue lines. Here, we note that Mo$_6$Te$_6$ wires appear at the very edge of the sample islands (c). (e)-(h) Atomic-resolution STM images of the marked regions in (a)-(d). Illustration of the atomic structure of T'-MoTe$_2$ (a), pristine H-MoTe$_2$ (b), and a Mo$_6$Te$_6$ bi-wire (d). Here, the MoTe$_{2-x}$ phases are formed through modification of line defects, *i.e.*, MTBs, which are chalcogen deficient (Mo:Te ratio of 1:1). (g) Typical MTB pattern of two bright parallel rows at a negative sample voltage (−1.0 V), with the pristine H-MoTe$_2$ pattern located on both sides. The insets of (e)-(h) show top-view schematics of the atomic models. The green and blue balls represent Te and Mo atoms, respectively. The unit cells are labelled with red polygonal shapes—rectangles for (e, h) and rhombuses for (f, g)—with their corresponding lattice constants detailed alongside. The MTB regions are highlighted by red shading (g). Scanning parameters: (a)-(d) bias voltage $V_{bias}$ = 2 V, tunnelling current $I_t$ = 10 pA; (e) 0.1 V, 100 pA; (f) −1.0 V, 50 pA; (g) −1.0 V, 100 pA; (h) 0.1 V, 100 pA.

Figure 1 shows a phase-temperature diagram of the ML Mo-Te compounds fabricated on BLG/SiC (0001) substrates (see Supplementary Material for details). By controlling the growth temperature ($T_{growth}$) of the substrate from 150 °C to 550 °C, we obtain four distinct phases: T'-MoTe$_2$[20], H-MoTe$_2$[20], multistoichiometric MoTe$_{2-x}$,[19-23] and 1D Mo$_6$Te$_6$ NWs.[11,24] Figures 1(a)-1(d) display large-scale STM topographic images of various phases. The insets are the line profiles taken along the blue lines, revealing the ML island height for each phase. Notably, wire-like features appear along the edges of the ML islands [as shown in Fig. 1(c)] as $T_{growth}$ reaches 350 °C and completely occupy the sample surface as the temperature is further increased [Fig. 1(d)], which are attributed to Mo$_6$Te$_6$ NWs. Close-up high-resolution STM images of these patterns are displayed in Figs. 1(e)-1(h), revealing the detailed atomic structure characteristics. Figure 1(e) shows the characteristic stripe structures of the 1T' phase with lattice constants of 0.63 nm and 0.35 nm, while Fig. 1(f) clearly displays a hexagonal symmetry pattern with a lattice constant of ~ 0.36 nm. Bi-wire arrays with lattice constants of 0.46 nm and 0.97 nm are depicted in Fig. 1(h). The above measurement results are in good agreement with those of previous studies.[20,24] The red rhombuses or rectangles indicate the unit cells of the various phases, with the corresponding top-view atomic models shown in the insets.

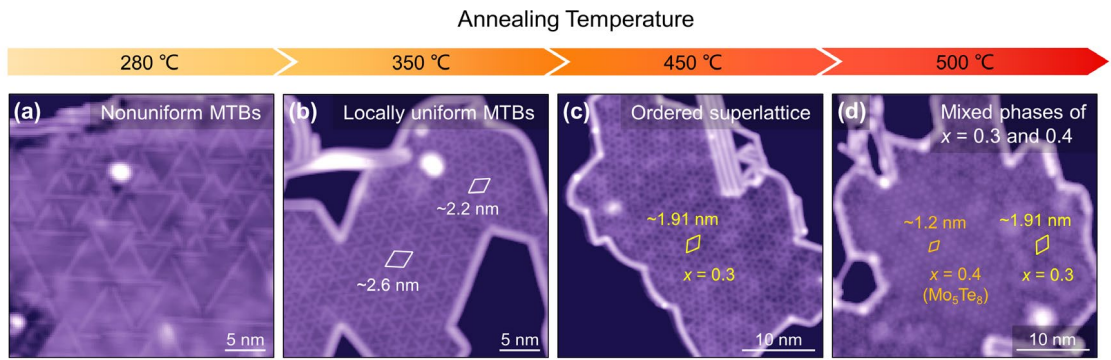

**Fig. 2.** Manipulation of the phase transformation in ML MoTe$_{2-x}$ via thermal annealing. (a)-(d) High-

resolution STM topographies of the sample annealing at different substrate temperatures, displaying a phase transition from nonuniform to uniform (with a definite $x$ value). (a) MTBs with triangular patterns that are nonuniform in size. (b) Locally ordered networks (indicated by white lines), typified by the "wagon-wheel" pattern as previously reported.[26-29] (c) Ordered MTB superlattice with a uniform periodicity of ~1.91 nm, *i.e.*, the MoTe$_{2-x}$ phase with $x = 0.3$ (marked by yellow lines). (d) Two distinguishable regions with different $x$ values: one possesses hexagonal symmetry with an ~1.2 nm superlattice periodicity, which corresponds to the MoTe$_{2-x}$ phase with $x = 0.4$, *i.e.*, Mo$_5$Te$_8$, as reported recently (marked by orange lines; more details see supplementary material, Fig. S1);[22,23] the other is identified as the MoTe$_{2-x}$ phase with $x = 0.3$, which is positioned on the periphery of the island. Here, we note that the Mo$_5$Te$_8$ phase tends to have a fragmented distribution over the sample surface, while the MoTe$_{2-x}$ phase with $x = 0.3$, can be obtained as the overwhelming majority over the entire ML sample (supplementary material, Figs. S2 and S3). (a)-(d) During the annealing process, Mo$_6$Te$_6$ wires appear at the very edge of the sample flakes. Scanning parameters: (a)-(c) bias voltage $V_{bias} = 2$ V, tunnelling current $I_t = 20$ pA; (d) 1 V, 20 pA.

The appearance of MTBs is usually inevitable in the synthesis of ML H-phase TMDs via bottom-up methods such as chemical vapour deposition (CVD) and MBE.[24-29] As one of the most common defects, an MTB is characterized by a chalcogen atom deficiency; hence, the stoichiometry is denoted MoTe$_{2-x}$. In our work, the typical MTB pattern, featuring two bright parallel rows at a negative sample voltage (–1.0 V), was identified, as shown by the red shadows highlighted in Fig. 1(g).[25] Generally, the formation of MTB triangular patterns of nonuniform sizes was observed in prior research.[24-29] However, through a controllable postgrowth annealing process, we effectively manipulated the evolutionary process of MTB triangular loops (*i.e.*, realized a change in the MTB triangular loops from local defects to global superstructures) to facilitate refined stoichiometry-dependent structural phase transformations in ML MoTe$_{2-x}$.

Here, we observed a special evolutionary process of the triangular patterns of MTBs by precisely controlling the annealing temperature ($T_{anneal}$) with the annealing time fixed at one hour, as depicted in Fig. 2. When $T_{anneal}$ is relatively low (280 °C), the sample surface displays notably nonuniform triangular patterns, with MTBs of random boundary lengths ranging in size from 0.7 to 13 nm [Fig. 2(a)]. When $T_{anneal}$ reaches 350 °C, locally uniform MTBs (or "wagon-wheel patterns") emerge [Fig. 2(b)].[26-29] The ~2.2 nm periodic structure observed within the domain was demonstrated to possess CDW order in recent studies.[19,20] With increasing temperature, the uniform superstructure adjusted via the triangular MTBs becomes discernible [Figs. 2(c) and 2(d)]. At $T_{anneal}$ = 450 °C, we obtain an ordered MTB superlattice with a uniform periodicity of 1.91 nm [marked by yellow lines; Figs. 2(c) and 2(d)]. Its stoichiometry, $MoTe_{2-x}$ with $x = 0.3$, is given by our latest work.[21] At $T_{anneal}$ = 500 °C, a distinct phase for $x = 0.4$, *i.e.*, $Mo_5Te_8$ (marked by orange lines), appears on the sample surface, exhibiting clear hexagonal arrays with an ~1.2 nm periodicity [Fig. 2(d); see supplementary material, Fig. S1 for identification].[22,23] The two phases show different distribution characteristics on the sample surface: The $Mo_5Te_8$ phase appears fragmented on the sample surface, while the $MoTe_{2-x}$ phase with $x = 0.3$ can be obtained as the overwhelming majority over the entire ML sample [Figs. 5(a) and 5(d); supplementary material, Figs. S2 and S3].

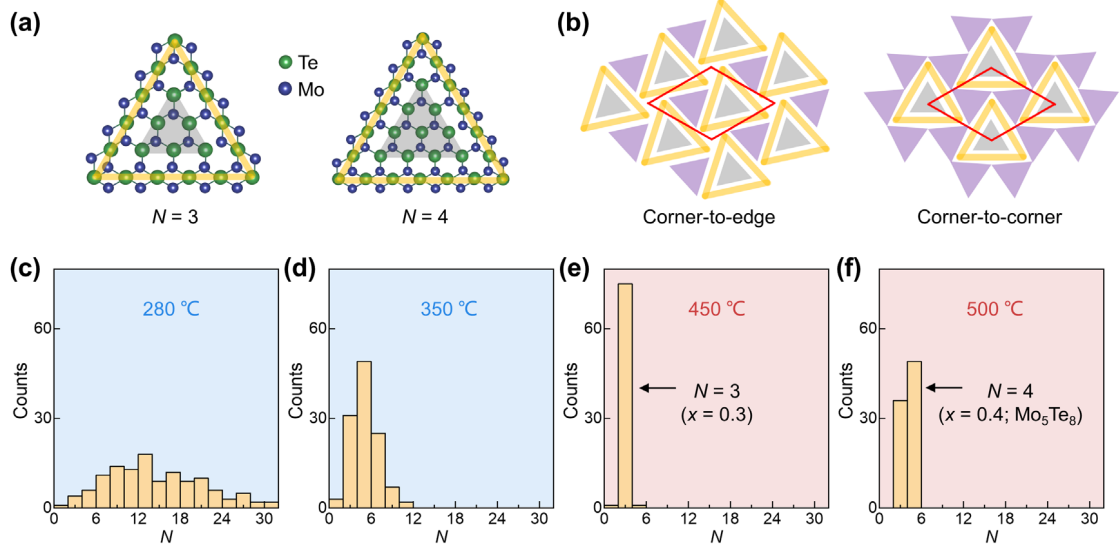

**Fig. 3.** (a) Illustration of the atomic models of the triangular domains enclosed by MTBs of different sizes at $N = 3$ and 4. Here, $N$ denotes the number of Te atoms at the domain (*i.e.*, grey triangle) edges. To guide the eye, we highlight the Te rows in the middle of the boundaries by orange lines. (b) Illustration of the MTB triangles arranged in different manners. There are two types of triangular domains: the grey domains are surrounded by MTB loops (orange triangles), whereas the other domains, indicated by purple triangles, are not. Left panel: the purple triangles are arranged in a corner-to-edge manner. Right panel: the purple triangles are connected in a corner-to-corner manner. In these two manners, the smallest periods occur for the MoTe$_{2-x}$ phases with $x = 0.3$ and $x = 0.4$ (Mo$_5$Te$_8$), corresponding to $N = 3$ and 4, respectively. The red rhombuses indicate the supercells. (c)-(f) Statistical analysis of the $N$ distribution at various annealing temperatures. When the temperature reaches 450 °C, $N = 3$ (*i.e.*, $x = 0.3$) dominates (e), but at 500 °C, $N = 4$ (*i.e.*, $x = 0.4$; Mo$_5$Te$_8$) slightly takes over.

Furthermore, we performed statistical measurements to clearly elucidate the phase evolution process. The atomic structures of the MTBs have been verified to consist of a Te atomic row sandwiched by two Mo atomic rows, as illustrated in the inset of Fig. 1(g).[26,27] Fig. 3a depicts the atomic models of two MTB triangles of sizes $N = 3$ and 4, where $N$ denotes the number of Te atoms at the domain (*i.e.*, grey triangle) edges.[21,22] To guide the eye, we highlight the Te rows in the middle of the boundaries by orange lines. Figure 3(b) illustrates the various arrangements of the MTB triangular domains, which include two types. The grey domains are encircled by MTB loops (shown as orange triangles), while the other type of domain (purple triangles) is not. In the left and right panels of Fig. 3(b), the purple triangular domains can be seen in corner-to-edge and corner-to-corner arrangements, respectively. Within these configurations, the smallest periods occur for the MoTe$_{2-x}$ phases with $x$ = 0.3 and $x$ = 0.4 (Mo$_5$Te$_8$), corresponding to $N$ = 3 and 4, respectively, with the red rhombuses denoting the supercell of the two lattices.[21,22] Figures 3(c)-3(f) show the statistical distribution of the triangular domain size (represented by $N$) for more than several hundred triangles in the different samples studied. We discover a random distribution of triangular domains, with $N$ ranging from 1 to 32, for $T_{anneal}$ = 280 °C [Fig. 3(c)]. At $T_{anneal}$ = 350 °C, the sizes of the triangular domains tend to shrink, *i.e.*, $N$ < 12 [Fig. 3(d)]. Upon reaching an annealing temperature of 450 °C, a prominent peak at $N$ = 3 emerges [Fig. 3(e)], signifying the formation of a thermally preferred phase—the MoTe$_{2-x}$ phase with $x$ = 0.3—possessing an ordered 1.91-nm superlattice. Finally, there are two dominant values of $N$ = 3 and 4 at $T_{anneal}$ = 500 °C [Fig. 3(f)], indicating the existence of two phases, namely, the MoTe$_{2-x}$ phases with $x$ = 0.3 and $x$ = 0.4 (Mo$_5$Te$_8$).

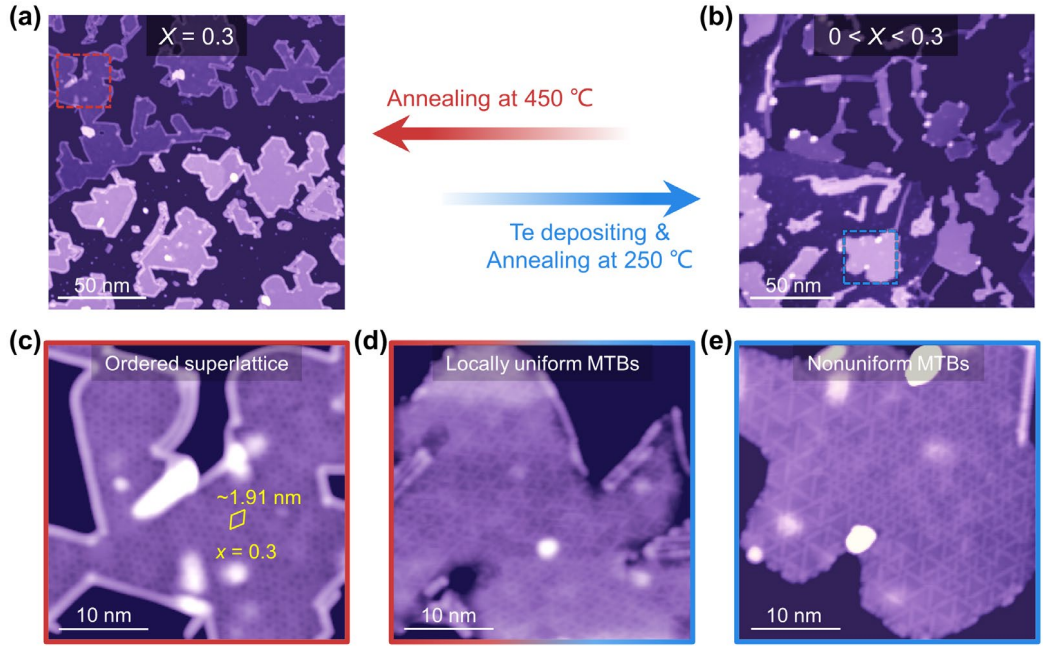

**Fig. 4.** Reversible phase transformation in ML MoTe$_{2-x}$. Large-scale STM topographic images of the uniform-sized MTB phase with $x = 0.3$ (a) and the nonuniform MTB phase with a range of $x$ values, *i.e.*, $0 < x < 0.3$ (b). The uniform phase can be achieved by annealing the nonuniform phase at 450 °C, whereas the nonuniform phase can be obtained by annealing the uniform phase at 250 °C in conjunction with Te deposition. (c, e) Atomic-resolution STM images of the marked regions indicated by the wine-red and blue dashed lines in (a) and (b), respectively. (c)-(e) Evolutionary process of the reversible transformation from the uniform phase to the nonuniform phase [contrary to what is illustrated in Figs. 2(a)-2(c)]. Scanning parameters: (a)-(e) $V_{bias} = 2$ V; (a, b) $I_t = 10$ pA; (c)-(e) $I_t = 30$ pA.

Moreover, our subsequent work demonstrates that the phase transformation process described above is not only feasible but also reversible. Although a few studies have mentioned the reversible transformation among various phases of Mo-Te compounds, such as Mo$_5$Te$_8$ ($x = 0.4$) to H-MoTe$_2$ and H-MoTe$_2$ to T′-MoTe$_2$,[23,30,31] a comprehensive depiction of the evolution process and regulation of the factors affecting the transformation remain lacking. We then concentrated on the multistoichiometric MoTe$_{2-x}$ structures to supplement and clarify this issue. Figure 4 presents a schematic representation of the experimental process for inducing the reversible phase transformation in ML MoTe$_{2-x}$. The large-scale and zoom-in atomic-resolution STM topographic images of the uniform/nonuniform phase are displayed in Figs. 4(a)/(b) and 4(c)/(d), respectively. Here, the uniform phase is the MoTe$_{2-x}$ phase with $x = 0.3$, while the nonuniform phase has a range of $x$ values, *i.e.*, $0 < x < 0.3$. The nonuniform phase can transform into the uniform phase with an ordered 1.91-nm superlattice when the annealing temperature reaches 450 °C (as indicated by the red arrow), as mentioned above. However, intriguingly, when the experimental conditions are altered to a low annealing temperature of $T_a = 250$ °C accompanied by deposition in a Te atmosphere (blue arrow), the previously visible uniform phase vanishes, yielding to the progressive dominance of nonuniform triangular domains proliferating across the sample surface, as illustrated in Figs. 4(c)-4(e).

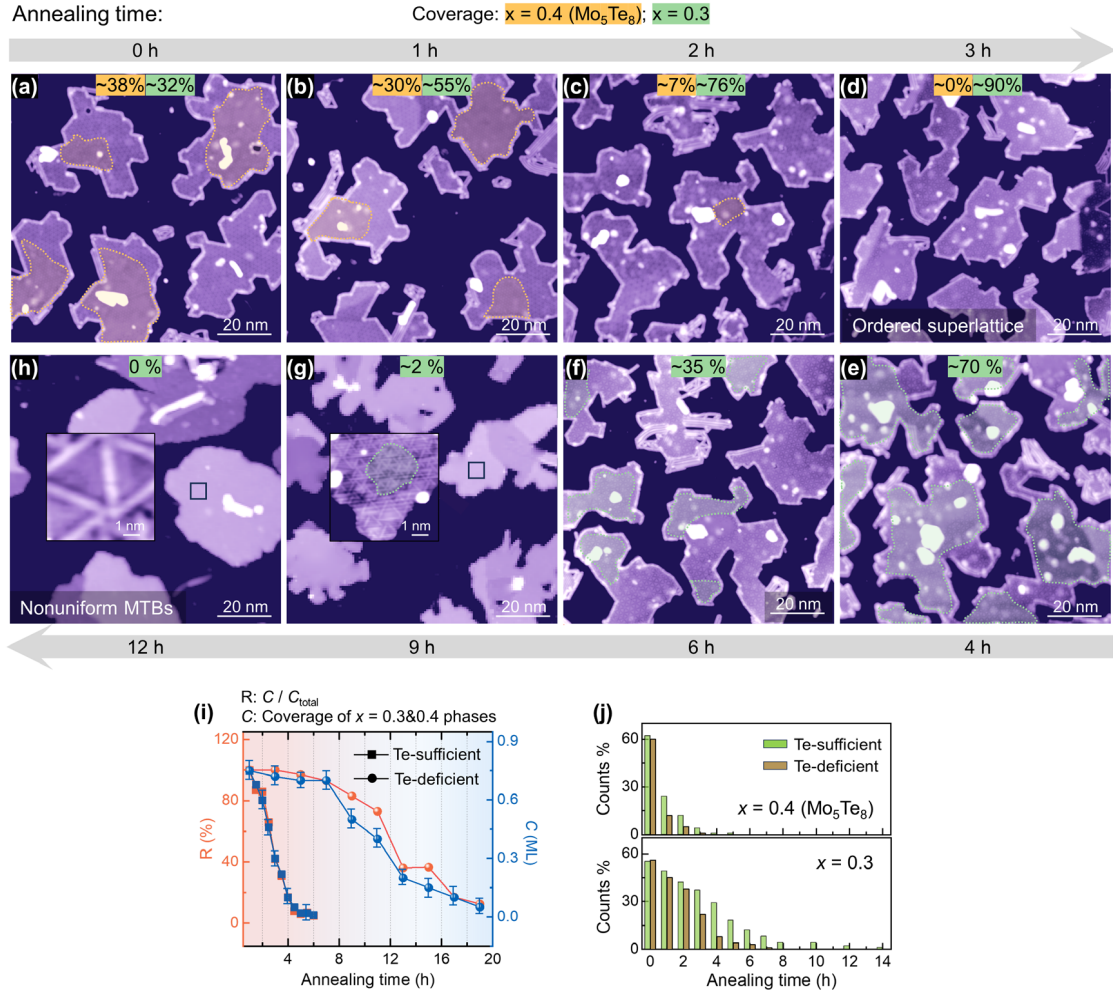

**Fig. 5.** Investigation of the annealing time and Te concentration effects on the phase transformation in MoTe$_{2-x}$. (a)-(h) Topographic evolution of ML MoTe$_{2-x}$ as a function of annealing time under Te-sufficient conditions. The regions of the two MoTe$_{2-x}$ phases with $x = 0.4$ (Mo$_5$Te$_8$) and $x = 0.3$ are distinctly framed by orange and green patches, respectively. When the annealing time increases to 3 hours, the $x = 0.3$ phase with an ordered superlattice emerges and dominates the entire sample surface (d). When the annealing time reaches 12 hours, nonuniform MTBs are identified in (h). The insets of (g, h) show close-up views of the STM topographic images for the regions indicated by black squares. (i) Plot of the coverage of the $x = 0.4$ and 0.3 phases ($C$) and coverage ratio R = $C/C_{total}$ as functions of annealing time. (j) Statistical analysis of the counts of the two phases: MoTe$_{2-x}$ with $x = 0.4$ (upper panel) and $x = 0.3$ (lower panel) with the annealing time under Te-sufficient and Te-deficient conditions. Scanning parameters: (a)-(h) $V_{bias} = 1$ V, $I_t = 20$ pA; the insets of (g, h) are $V_{bias} = 2$ V, $I_t = 10$ pA.

Apart from the notable influence of the annealing temperature on the phase formation and crystal structures in the as-grown 2D crystals, we further investigated the other two growth parameters that influence the phase transformation: the annealing time and Te concentration. Figures 5(a)-5(h) depict a series of high-resolution STM images showing the evolution of the ML MoTe$_{2-x}$ surface topography as a function of annealing time under Te-sufficient conditions. Here, we take the two MoTe$_{2-x}$ phases with $x = 0.4$ (Mo$_5$Te$_8$) and $x = 0.3$ as references. The regions of the two phases are distinctly framed by orange ($x = 0.4$) and green ($x = 0.3$) patches. Initially, both phases have similar coverage, as depicted in Fig. 5(a). However, the $x = 0.4$ phase rapidly disappears with increasing annealing time, and the $x = 0.3$ phase coverage becomes predominant [Fig. 5(d)]. As the time further increases, the $x = 0.3$ phase also gradually disappears, and eventually, the entire sample surface is completely covered by relatively large MTB triangles, which are close to pristine MoTe$_2$ ($x = 0$). We measured numerous samples, and the results are illustrated in Fig. 5(i). This Fig. displays the coverage of the two mixed phases with $x = 0.3$ and $x = 0.4$ (C) and the coverage ratio R = $C/C_{total}$ as functions of annealing time under both Te-sufficient and Te-deficient conditions. The statistical analysis of the counts of the $x = 0.4$ (upper panel) and $x = 0.3$ (lower panel) phases is shown in Fig. 5(j). In our experiment, Te-sufficient conditions were reached by directly depositing 2 nm of Te before annealing, while Te-deficient conditions were achieved by depositing Te at a rate of 0.4 nm/min during the annealing process. The counts of the mixed phases gradually decrease as the annealing time increases, giving way to a surge in the nonuniform MTB triangles and a quicker transition from the uniform to nonuniform phase under Te-sufficient conditions. Notably, the Mo$_5$Te$_8$ phase disappears over a shorter annealing time compared to the MoTe$_{2-x}$ phase with $x = 0.3$ (supplementary material, Fig. S4). These results further reinforce our conclusions about how the annealing time and Te concentration considerably influence the phase transformations.

In summary, our work systematically investigated the multistage transformation in ML Mo-Te compounds. We conducted a comprehensive multiparameter tuning study and revealed the temperature-dependent phase diagram from T'-MoTe to H-MoTe$_2$, MoTe$_{2-x}$ ($x$ = 0 to 0.4) and 1D Mo$_6$Te$_6$ NWs, which clarifies the confusion regarding the complex structural phases of Mo-Te observed in previous studies. Additionally, we found that MoTe$_{2-x}$ with $x$ = 0.3 can be obtained as an isolated and globally formed phase in MBE growth. An important aspect of our work is achieving controllably reversible transformation among these multistoichiometric phases, which significantly contributes to our current understanding of the phase evolution. Overall, the discovery of the crucial role of growth parameters not only provides an essential understanding of emerging 2D Mo-Te materials but also offers a systematic strategy for manipulating phase transformations. The detailed insights into the polymorphic forms of Mo-Te films reinforce the capacity of these materials to act as a platform for exploring quantum physics and inspire potential applications in future phase-controlled electronic and optoelectronic devices.

## SUPPLEMENTARY MATERIAL

See the supplementary material for more details of sample preparation and STM measurements, the identification of the $Mo_5Te_8$ phase (Fig. S1), the different distribution characteristics on the sample surface of the $Mo_5Te_8$ phase (Fig. S2) and the $MoTe_{2-x}$ phase with $x = 0.3$ (Fig. S3), and the influence of the Te concentration on the phase transition (Fig. S4).

This work was supported by the National Natural Science Foundation of China (12134011), the National Key R&D Program of China (Grant No. 2018YFA0703700).

## AUTHOR DECLARATIONS

### Conflict of Interest

The authors have no conflicts to disclose.

**Author Contributions:**

**Zemin Pan:** Conceptualization (equal); Formal analysis (lead); Investigation (lead); Methodology (lead); Writing – original draft (lead); Writing – review & editing (equal). **Tao Jian:** Data curation (equal); Formal analysis (equal). **Hui Zhang:** Data curation (equal); Investigation (equal). **Xiaoyu Lin:** Investigation (equal). **Chao Zhu:** Investigation (equal). **Jinghao Deng:** Methodology (equal). **Zhengbo Cheng:** Methodology (equal). **Chuansheng Liu:** Funding acquisition (equal); Supervision (equal). **Chendong Zhang:** Conceptualization (equal); Funding acquisition (lead); Investigation (equal); Supervision (equal); Writing – original draft (equal); Writing – review & editing (equal).

## DATA AVAILABILITY

The data that support the findings of this study are available from the corresponding author upon reasonable request.